\documentclass[12pt]{article}
\begin{document}
\begin{center}
{\large\bf M-Theory and de Sitter Space}
\vskip 0.3 true in
{\large J. W. Moffat}
\vskip 0.3 true in
{\it Department of Physics, University of Toronto,
Toronto, Ontario M5S 1A7, Canada}
\end{center}
\begin{abstract}%
An M-theory constructed in an eleven-dimensional supermanifold with
a $\diamondsuit$-product of field operators is shown to have a de Sitter
space solution. Possible implications of this result for cosmology are
mentioned.
\end{abstract}
\vskip 0.2 true in e-mail:
moffat@medb.physics.utoronto.ca


\section{Introduction}

Recently, an M-theory has been formulated in an eleven-dimensional
supermanifold with coordinates~\cite{Moffat}
\begin{equation}
\rho^M=x^M+\beta^M,
\end{equation}
where $M,N=0,1,...10$ and $x^M$ are commuting coordinates
and $\beta^M$ are Grassmann anticommuting coordinates.
Both noncommutative and non-anticommutative geometries can be unified
within the superspace formalism using the $\circ$-product of two
operators ${\hat f}$ and ${\hat g}$~\cite{Moffat2,Moffat3,Moffat4}:
\begin{equation}
\label{fgproduct}
({\hat f}\circ{\hat g})(\rho)
=\biggl[\exp\biggl(\frac{1}{2}\omega^{MN}
\frac{\partial}{\partial\rho^M}\frac{\partial}{\partial\eta^N}\biggr)
f(\rho)g(\eta)\biggr]_{\rho=\eta} $$ $$
=f(\rho)g(\rho)+\frac{1}{2}\omega^{MN}\partial_M
f(\rho)\partial_Ng(\rho)+O(\vert\omega\vert^2),
\end{equation}
where $\partial_M=\partial/\partial\rho^M$ and $\omega^{MN}$ is a
nonsymmetric tensor
\begin{equation}
\omega^{MN}=\tau^{MN}+i\theta^{MN},
\end{equation} with
$\tau^{MN}=\tau^{NM}$ and $\theta^{MN}=-\theta^{NM}$.
Moreover, $\omega^{MN}$ is Hermitian symmetric
$\omega^{MN}=\omega^{\dagger MN}$, where $\dagger$ denotes Hermitian
conjugation. The familiar commutative coordinates of spacetime are replaced
by the superspace operator relations
\begin{equation}
\label{commutator}
[{\hat\rho}^M,{\hat\rho}^N]=2\beta^M\beta^N+i\theta^{MN},
\end{equation}
\begin{equation}
\label{anticommutator}
\{{\hat\rho}^M,{\hat\rho}^N\}=2x^Mx^N+2(x^M\beta^N+x^N\beta^M)+\tau^{MN}.
\end{equation}
In the limits that $\beta^M\rightarrow 0$ and
$\vert\tau^{MN}\vert\rightarrow 0$, we get the familiar expression for
noncommutative coordinate operators
\begin{equation}
[{\hat x}^M,{\hat x}^N]=i\theta^{MN}.
\end{equation}
In the limits $x^M\rightarrow 0$ and $\vert\theta^{MN}\vert\rightarrow 0$,
we obtain the Clifford algebra anticommutation relation
\begin{equation}
\{{\hat\beta}^M,{\hat\beta}^N\}=\tau^{MN}.
\end{equation}

We used the simpler non-anticommutative
geometry obtained when $\theta^{MN}=0$ to construct the M-theory,
because it alone can lead to a finite and unitary quantum field theory and
quantum gravity theory ~\cite{Moffat2,Moffat3,Moffat4}.
In the non-anticommutative field theory formalism, the product of two
operators ${\hat f}$ and ${\hat g}$ has a corresponding
$\diamondsuit$-product
\begin{equation}
\label{diamondproduct}
({\hat f}\diamondsuit {\hat g})(\rho)
=\biggl[\exp\biggr(\frac{1}{2}\tau^{MN}\frac{\partial}{\partial\rho^M}\frac{\partial}
{\partial\eta^N}\biggr)f(\rho)g(\eta)\biggr]_{\rho=\eta} $$ $$
=f(\rho)g(\rho)+\frac{1}{2}\tau^{MN}\partial_M f(\rho)\partial_N
g(\rho)+O(\tau^2).
\end{equation}
We chose as the basic action of the
M-theory, the Cremmer, Julia and Scherk (CJS)~\cite{Julia} action for
eleven-dimensional supergravity, replacing all products of field operators
with the $\diamondsuit$-product. This action is invariant under the
generalized $\diamondsuit$-product supersymmetry gauge transformations of
the vielbein $e^A_M$, the spin 3/2 field $\psi_M$ and the three-form field
$A_{MNQ}$. Our M-theory has as a possible low energy limit the CJS
supergravity theory when $\vert\tau^{MN}\vert\rightarrow 0$ and
$\beta^M\rightarrow 0$.

In previous work~\cite{Moffat2,Moffat3,Moffat4}, we
demonstrated that scalar quantum field theory and weak field quantum
gravity are finite to all orders of perturbation theory, and the S-matrix
for these theories is expected to obey unitarity. An analysis of the
higher-dimensional field theories, including supersymmetric gauge
theories generalized to the non-anticommutative formalism, shows that they
will also be finite to all orders of perturbation theory. This result is
mainly due to the existence of a fundamental length scale $\ell$ in the
theory that owes its existence to a quantization of spacetime. When
$\ell\rightarrow 0$ the non-anticommutative field theories reduce to the
standard local point particle field theories which suffer the usual
difficulties of infinities and non-renormalizable quantum gravity. The
modification of standard local point field theory occurs at short
distances or high energies where accelerator physics has not been probed.

There is now considerable observational evidence for the acceleration of
the expansion of the universe~\cite{Perlmutter}. This means that dark
energy is the predominant form of matter in the present universe, which
can be described either by quintessence~\cite{Caldwell} or a positive
cosmological constant. The present data can be fitted with $\Omega_M\sim
0.3$ and $\Omega_{\lambda}\sim 0.7$, where $\Omega_M$ and
$\Omega_{\lambda}$ are the ratios of the matter density $\rho_M$ and the
vacuum density $\rho_{\rm vac}$ to the
critical density $\rho_c$, respectively. Hopefully, future supernovae
observational data will be able to distinguish between a universe with a
positive cosmological constant and quintessence~\cite{Copeland}. In any
event, this requires that our low energy M-theory has a four-dimensional de
Sitter space solution.

In general, it is difficult to obtain a cosmological
solution with a horizon from perturbative superstring theory and
supergravity theory~\cite{Witten}. There exist no-go theorems preventing
the existence of solutions in eleven-dimensional supergravity with a
non-vanishing cosmological constant~\cite{Maldacena}. A spontaneous
compactification of this theory leads to an anti-deSitter (3+1) spacetime
and a compact $S^7$ space with a positive cosmological constant. Demanding
a supersymmetric vacuum leads to an anti-deSitter solution for spacetime,
which cannot explain the current cosmological observations. In
~\cite{Moffat}, we extended our generalized M-theory to the massive
M-theory formulation of Chamblin and Lambert~\cite{Lambert}, which reduced
to a massive type-IIA supergravity theory in ten dimensions. This scheme
was able to produce a de Sitter spacetime solution in four-dimensional
spacetime by a compactification of the ten-dimensional theory. However,
potential problems of stability of the theory exist and there is no action
for the field equations~\cite{Lambert2}.

In the following, we shall investigate another approach to the problem
of obtaining de Sitter space solutions of our M-theory field
equations by using the Freund-Rubin~\cite{Freund,Freund2} ansatz for the
four-form field $F_{MNPQ}$, neglecting the fermion contributions, and
expanding the $\diamondsuit$-product of the F-fields. We shall find that we
can obtain a de Sitter (3+1) spacetime solution with a positive
cosmological constant.

\section{\bf M-theory Bosonic Field Equations}

The bosonic action of the M-theory takes the form
\begin{equation}
S=\int
d^{(11)}\rho\sqrt{g^{(11)}}\diamondsuit\biggl[-\frac{1}{2}R-\frac{1}{48}F_{MNPQ}\diamondsuit
F^{MNPQ}+\biggl[\frac{\sqrt{2}}{6\cdot(4!)^2}\biggr]\biggl(\frac{1}{\sqrt{g^{(11)}}}\biggr)
$$ $$ \times\diamondsuit
\epsilon^{M_1M_2...M_{11}}F_{M_1M_2M_3M_4}\diamondsuit
F_{M_5M_6M_7M_8}\diamondsuit A_{M_9M_{10}M_{11}}\biggr].
\end{equation}
The metric is $(-+++...+)$, $\epsilon^{0123...}=+1$ and
$F_{MNPQ}=4!\partial_{[M}A_{NPQ]}$ and we have set $8\pi G^{(11)}=c=1$,
where $G^{(11)}$ is the eleven-dimensional gravitational coupling constant.

The field equations are
\begin{equation}
R_{MN}-\frac{1}{2}g_{MN}\diamondsuit R=-T_{F\,MN},
\end{equation}
\begin{equation}
\label{Fequation}
\frac{1}{\sqrt{g^{(11)}}}\diamondsuit\partial_M(\sqrt{g^{(11)}}\diamondsuit
F^{MNPQ})
=-\biggl[\frac{\sqrt{2}}{2\cdot(4!)^2}\biggr]\biggl(\frac{1}{\sqrt{g^{(11)}}}\biggr)
\diamondsuit\epsilon^{M_1...M_8NPQ}
$$ $$ \times
F_{M_1M_2M_3M_4}\diamondsuit
F_{M_5M_6M_7M_8},
\end{equation}
where
\begin{equation}
\label{Ftensor}
T_{F\,MN}=\frac{1}{48}(8F_{MPQR}\diamondsuit {F_N}^{PQR}-g_{MN}\diamondsuit
F_{SPQR}\diamondsuit F^{SPQR}).
\end{equation}

\section{Cosmological Solutions}

For cosmological purposes, we shall restrict our attention to an
eleven-dimensional metric of the form
\begin{equation}
g_{MN}=\left(\begin{array}{ccc}
        -1&0&0\\
        0 & a_4^2(t){\tilde g}_{ij}&0\\
        0 & 0 & a_7^2(t){\tilde {g}}_{\Delta\Sigma}
        \end{array}\right).
\end{equation}
Here, $\tilde{g}_{ij}$ (i,j =1,2,3) and $\tilde{g}_{\Delta\Sigma}$
$(\Delta\Sigma=5,...,11)$ are the maximally symmetric three and seven
spacelike spaces, respectively, and $a_4$ and $a_7$ are the corresponding
time dependent cosmological scale factors. We have assumed for simplicity
that the seven extra spacelike dimensions form a maximally symmetric space,
although there is no a priori reason that this be the case. We shall assume
that the Grassmann coordinates $\beta^M$ are small compared to $x^M$,
$\rho^M\approx x^M$.

The non-vanishing components of the Christoffel symbols are
\begin{equation}
\Gamma^0_{ij}=\frac{{\dot a}_4}{a_4}g_{ij},\quad
\Gamma^0_{\Delta\Sigma}=\frac{{\dot a}_7}{a_7}g_{\Delta\Sigma},\quad
\Gamma^i_{j0}=\frac{{\dot a}_4}{a_4}{\delta^i}_j,
$$ $$
\Gamma^\Gamma_{\Delta 0}=\frac{{\dot
a}_7}{a_7}{\delta^\Gamma}_\Delta,\quad
\Gamma^i_{jk}={\tilde\Gamma}^i_{jk},\quad
\Gamma^\Gamma_{\Delta\Sigma}={\tilde\Gamma}^\Gamma_{\Delta\Sigma},
\end{equation} where $g_{ij}=a^2_4{\tilde g}_{ij}$,
$g_{\Delta\Sigma}=a^2_7{\tilde g}_{\Delta\Sigma}$ and
${\tilde\Gamma}^i_{jk}$ and ${\tilde\Gamma}^\Gamma_{\Delta\Sigma}$ are the
Christoffel symbols formed from the ${\tilde g}_{ij}$, ${\tilde
g}_{\Delta\Sigma}$ and their derivatives.

The non-vanishing components of
the Ricci tensor are
\begin{equation}
\label{Ricci}
R_{00}=3\frac{{\ddot a_4}}{a_4}+7\frac{{\ddot a_7}}{a_7},
$$ $$
R_{ij}=-\biggl[\frac{2k_4}{a_4^2}
+\frac{d}{dt}\biggl(\frac{{\dot a}_4}{a_4}\biggr)+\biggl(3\frac{{\dot
a}_4}{a_4}+7\frac{{\dot a}_7}{a_7}\biggr)\frac{{\dot
a}_4}{a_4}\biggr]g_{ij},
$$ $$
R_{\Delta\Sigma}=-\biggl[\frac{2k_7}{a^2_7}+\frac{d}{dt}\biggl(\frac{{\dot
a}_7}{a_7}\biggr)+\biggl(3\frac{{\dot a}_4}{a_4}+7\frac{{\dot
a}_7}{a_7}\biggr)\frac{{\dot a}_7}{a_7}\biggr]g_{\Delta\Sigma},
\end{equation}
where $k_4$ and $k_7$ are the curvature constants of
four-dimensional and seven-dimensional space. Positive and negative values
of $k_4$ and $k_7$ correspond to the sphere and the pseudosphere,
respectively, while vanishing values of $k_4$ and $k_7$ correspond to flat
spaces.

We now adopt the Freund-Rubin ansatz for which all components of the
four-form field $F_{MNPQ}$ vanish except~\cite{Freund,Freund2}:
\begin{equation}
F_{\mu\nu\rho\sigma}=mf(t)\frac{1}{\sqrt{-g^{(4)}}}\epsilon_{\mu\nu\rho\sigma},
\end{equation}
where $\mu,\nu=0,1,2,3$ and $m$ is a constant. With this ansatz, the
trilinear contributions in $A_{MNP}$ and its derivatives in the action
vanish. We shall use (\ref{diamondproduct}) to expand the products of the
F-tensors in small values of $\vert\tau^{MN}\vert$. We shall neglect the
contributions from the $\diamondsuit$-products of the metric $g_{MN}$ and
its derivatives compared to the $\diamondsuit$-products of the F-tensors.
Using the results that
$\epsilon_{\mu\alpha\rho\sigma}{\epsilon_\nu}^{\alpha\rho\sigma}
=6m^2f^2(t)g_{\mu\nu}(-g^{(4)})$ and
$\epsilon_{\mu\nu\rho\sigma}\epsilon^{\mu\nu\rho\sigma}=24m^2f^2(t)(-g^{(4)})$,
we find to first order in $\vert\tau\vert$:
\begin{equation}
T_{F\,MN}=\frac{1}{2}\epsilon
m^2\biggl(f^2-\frac{{\dot f}^2}{2\Lambda^2}\biggr)g_{MN},
\end{equation}
where we have chosen
\begin{equation}
\tau^{00}=-\frac{1}{\Lambda^2},
\end{equation}
and $\Lambda$ is a constant with the dimensions of energy.
Moreover, $\epsilon =+1$ for $M,N=\mu,\nu$, and $-1$ for
$M,N=\Delta,\Sigma$. Eq.(\ref{Fequation}) becomes
\begin{equation}
\frac{d}{dt}\biggl[(a_7(t))^7f(t)\biggr]=0.
\end{equation}

We have
\begin{equation}
R_{MN}-\frac{1}{2}g_{MN}R=\lambda_{11} g_{MN},
\end{equation}
where $\lambda_{11}$ is the eleven-dimensional cosmological constant:
\begin{equation}
\lambda_{11}= -\frac{1}{2}\epsilon m^2\biggl(f^2-\frac{{\dot
f}^2}{2\Lambda^2}\biggr).
\end{equation}
This leads to
\begin{equation}
\label{R1} R_{\mu\nu}-\frac{1}{2}g_{\mu\nu}R=\lambda_4g_{\mu\nu},
\end{equation} and
\begin{equation}
\label{R2}
R_{\Delta\Sigma}-\frac{1}{2}g_{\Delta\Sigma}R=\lambda_7g_{\Delta\Sigma},
\end{equation}
where
\begin{equation}
\lambda_4=-\frac{1}{2}m^2\biggl(f^2-\frac{{\dot
f}^2}{2\Lambda^2}\biggr),\quad \lambda_7
=\frac{1}{2}m^2\biggl(f^2-\frac{{\dot f}^2}{2\Lambda^2}\biggr).
\end{equation}

Let us assume that
\begin{equation}
\label{Cconstant}
C\approx f^2-\frac{{\dot f}^2}{2\Lambda^2},
\end{equation}
where $C$ is a constant. In the
limit, $\Lambda\rightarrow\infty$, we obtain the standard supersymmetric
vacuum result
\begin{equation}
\lambda_4=-\frac{1}{2}m^2f^2,\quad \lambda_7=\frac{1}{2}m^2f^2.
\end{equation}
The eleven-dimensional space becomes a product of a four-dimensional
Einstein anti-de Sitter space with negative
cosmological constant and a seven-dimensional Einstein space with positive
cosmological constant~\cite{Duff}.
If we require the vacuum to be supersymmetric by
demanding covariantly constant spinors $\theta$ for which
\begin{equation}
\delta\psi={\bar D}_M\theta=0,
\end{equation}
where
\begin{equation}
{\bar D}_M=D_M+\frac{i\sqrt{2}}{288}(\Gamma^{NPQR}_M-8\Gamma^{PQR}
{\delta_M}^N)F_{NPQR},
\end{equation}
then for $m\not= 0$, $N=8$ supersymmetry uniquely chooses $AdS\times S^7$
with an $SO(8)$-invariant metric on $S^7$~\cite{Duff,Englert}.

If we choose $f^2={\dot f}^2/2\Lambda^2$, we
get $\lambda_4=\lambda_7=0$ and flat $(3+1)$ and
seven-dimensional spaces. On the other hand, if we choose
$f^2 < {\dot f^2}/2\Lambda^2$, then $\lambda_4 > 0$ and $\lambda_7 < 0$ and
we obtain a positive cosmological constant in $(3+1)$ spacetime,
corresponding to a de Sitter universe, and a seven-dimensional anti-de
Sitter space.

From (\ref{R1}) and (\ref{R2}), we obtain
\begin{equation}
\label{a1}
3{\ddot a}_4+7\frac{{\ddot a}_7a^2_4}{a_7}=\lambda_4a^2_4,
\end{equation}
\begin{equation}
\label{a2}
2k_4+{\ddot a}_4a_4+2{\dot a}^2_4+7\frac{{\dot a}_7{\dot
a}_4a_4}{a_7}=\lambda_4a^2_4,
\end{equation}
\begin{equation}
2k_7+{\ddot a}_7a_7+6{\dot a}^2_7
+3\frac{{\dot a}_4{\dot a}_7a_7}{a_4}=\lambda_7a^2_7.
\end{equation}
Combining (\ref{a1}) and (\ref{a2}) gives
\begin{equation}
\biggl(\frac{{\dot a}_4}{a_4}\biggr)^2+\frac{k_4}{a^2_4}
-\frac{7}{6}\frac{{\ddot a}_7}{a_7}
+\frac{7}{2}\frac{{\dot a}_7{\dot a}_4}{a_4a_7}=\frac{1}{3}\lambda_4.
\end{equation}
For solutions in which the scale factor $a_4$ expands
faster than $a_7$, we get the standard four-dimensional Friedmann equation:
\begin{equation}
\biggl(\frac{{\dot a}_4}{a_4}\biggr)^2
+\frac{k_4}{a^2_4}=\frac{1}{3}\lambda_4.
\end{equation}
This has the de Sitter inflationary solution for $k_4=0$ and $\lambda_4 >
0$:
\begin{equation}
a_4=B\exp\biggl(\sqrt{\frac{\lambda_4}{3}}t\biggr),
\end{equation}
where $B$ is a constant.

\section{Conclusions}

Our M-theory describes a finite quantum field theory in an
eleven-dimensional super manifold in which the action is constructed from a
$\diamondsuit$-product of field operators based on eleven-dimensional CJS
supergravity theory. The quantum gravity and quantum gauge field parts of
the action will be finite for $\Lambda < \infty$ due to the finiteness of
the non-anticommutative quantum field theory. In the limits
$\Lambda\rightarrow\infty$ and $\beta^M\rightarrow 0$, we obtain the low
energy limit of eleven-dimensional CJS supergravity. The compactified
version of this theory has the same massless ten-dimensional particle
spectrum as type-IIA superstring theory and is connected to the latter
theory by a duality transformation.

The M-theory eleven-dimensional field equations are invariant under
generalized $\diamondsuit$-product supersymmetric gauge transformations,
which can be thought of as classical deformations of the standard
supersymmetric gauge transformations of CJS supergravity. The generalized
$\diamond$-product supersymmetric vacuum leads to de Sitter space
solutions and thus provides an empirical basis for a realistic cosmology.
The no-go theorems in~\cite{Maldacena} do not apply in this case, because
they are derived from standard supersymmetric theories. Our M-theory with
generalized supersymmetric equations predicts an inflationary period in the
early universe. As the universe expands, $f^2$ can tend towards ${\dot
f}^2/2\Lambda^2$ and produce a small four-dimensional cosmological constant
with $\lambda_4 > 0$ and an accelerating universe at present. However, we
have neglected the fermion contributions to the generalized M-theory field
equations, so we have to include them and matter and radiation
contributions to the equations of motion to describe a fit to observational
data. This will be the subject of a future investigation.

\vskip 0.2 true in
{\bf Acknowledgments}
\vskip 0.2 true in
I thank Michael Clayton for helpful and
stimulating conversations. This work was supported by
the Natural Sciences and Engineering Research Council of Canada.
\vskip 0.5 true in

\end{document}